\documentclass[preprint,prl,showpacs,superscriptaddress]{revtex4}

\usepackage{graphicx}
\usepackage{bm}
\usepackage{amsmath}
\usepackage{amsfonts}

\newcommand{\beq}{\begin{equation}}
\newcommand{\eeq}{\end{equation}}
\newcommand{\beqa}{\begin{eqnarray}}
\newcommand{\eeqa}{\end{eqnarray}}
\newcommand{\vep}{\varepsilon}

\begin{document}


\title{Macroscopic quantum state in a semiconductor device}
 \author{Yun-Pil Shim}
 \altaffiliation{Present address: Department of Physics, University of Wisconsin-Madison, WI 53706}
 \affiliation{Institute for Microstructural Sciences, National Research Council of Canada,
              Ottawa, Canada K1A 0R6}
 \author{Anand Sharma}
 \affiliation{Institute for Microstructural Sciences, National Research Council of Canada,
              Ottawa, Canada K1A 0R6}
 \affiliation{Department of Physics, University of Ottawa, Ottawa, Canada K1N 6N5}
 \author{Chang-Yu Hsieh}
 \affiliation{Institute for Microstructural Sciences, National Research Council of Canada,
              Ottawa, Canada K1A 0R6}
 \affiliation{Department of Physics, University of Ottawa, Ottawa, Canada K1N 6N5}
 \author{Pawel Hawrylak}
 \affiliation{Institute for Microstructural Sciences, National Research Council of Canada,
              Ottawa, Canada K1A 0R6}

 \date{\today}

\begin{abstract}
We show how nanostructuring of a metallic gate on a field-effect transistor (FET) can lead to a macroscopic, 
robust and voltage controlled quantum state in the electron channel of a FET. 
A chain of triple quantum dot molecules created by gate structure realizes a spin-half Heisenberg chain 
with spin-spin interactions alternating between ferromagnetic and anti-ferromagnetic. 
The quantum state is a semiconductor implementation of an integer spin-one antiferromagnetic Heisenberg chain 
with a unique correlated ground state and a finite energy gap, originally conjectured by Haldane.
\end{abstract}

\maketitle

\begin{figure}
 \includegraphics[width=0.8\linewidth]{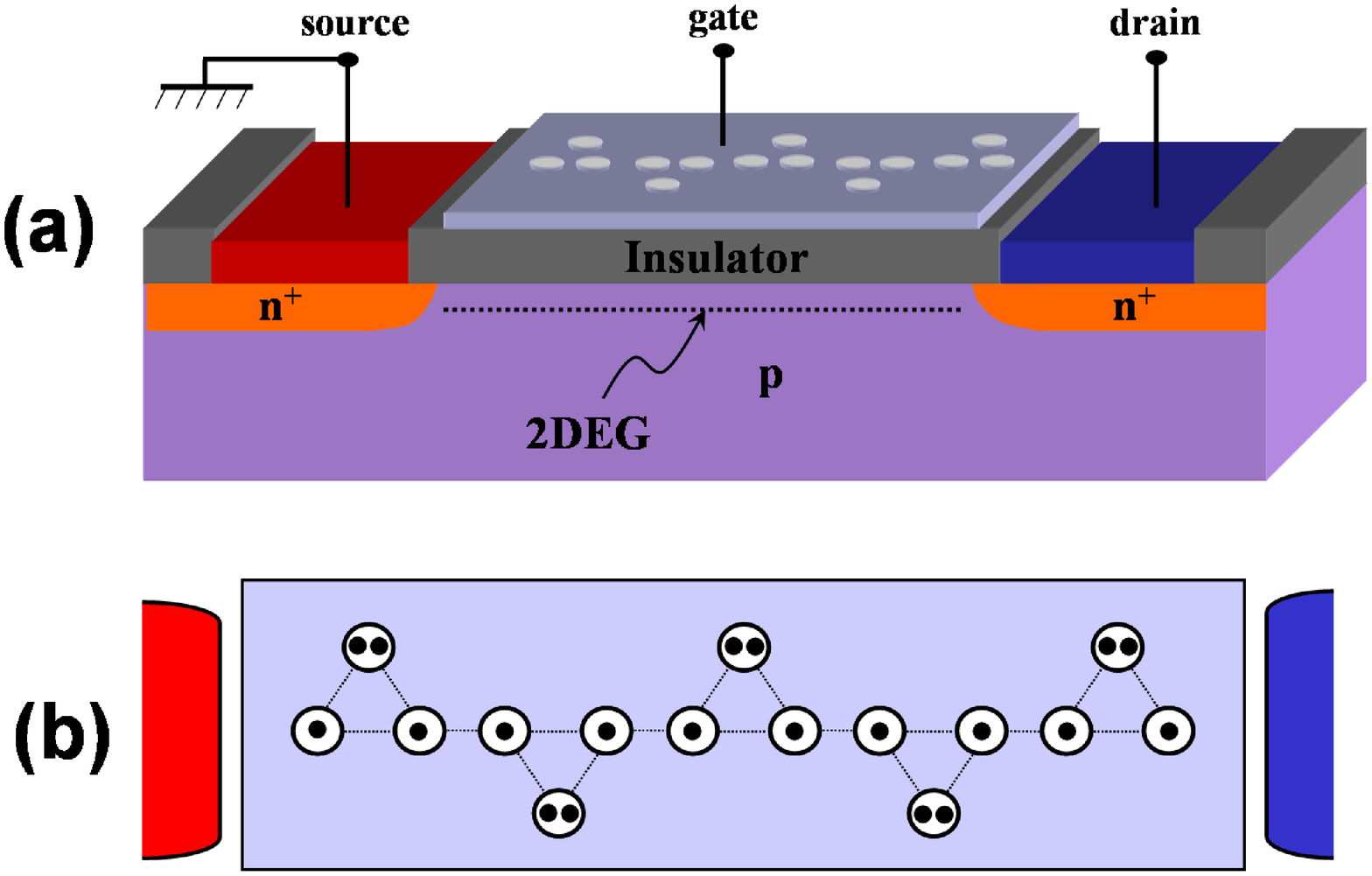}\\
 \caption{(Color online) Schematic picture of the triple quantum dot chain on a semiconductor device. 
         (a) The triple quantum dot chain is defined by a top gate, with a pattern of holes to create potential minima 
         in the two-dimensional electron gas in the active region below. 
         (b) For the realization of effective spin-one chain, each dot on the linear chain contains a single electron 
         and the other dots on the top and bottom contain two electrons.}
 \label{fig:setting}
\end{figure}

Haldane predicted that spin-one antiferromagnetic Heisenberg chain has a unique ground state with a finite gap \cite{haldane_physlett1983}.
Affleck {\it et al.} obtained an exact analytical solution of the ground state 
with a finite gap in an isotropic spin-one chain with a special biquadratic interaction \cite{affleck_kennedy_prl1987}
and suggested that the ground state of the Heisenberg spin-one chain has the valence-bond-solid state character \cite{affleck_kennedy_commun_math_phys1988}. 
These predictions were subsequently confirmed by numerical calculations \cite{nightingale_blote_prb1986,white_prl1992,white_huse_prb1993,white_affleck_prb2008} 
and experimental studies \cite{buyers_morra_prb1986,morra_buyers_prb1988,cizmar_ozerov_njp2008} 
in quasi-one-dimensional complex compounds. 
In this work, we propose a way of realizing an effective spin-one chain in a system of lateral gated quantum dots (QD) 
in a field effect transistor (FET) semiconductor structure. 
Quantum dots \cite{jacak_hawrylak_book1998} are defined by confinement potentials 
in all three dimensions and often called artificial atoms 
due to their similarity to the real atoms. Many QDs can be connected in a network to form more complex artificial structures. 
In a double quantum dot molecule, the ground state of two-electron system is always spin singlet without external magnetic field. 
Therefore, the effective spin-spin interaction between two localized electrons in respective quantum dots is always antiferromagnetic. 
It was shown by some of us that the spin-spin interaction can be tuned, both in magnitude and sign, 
by bringing another quantum dot that contains two electrons to make a triple quantum dot (TQD) molecule \cite{shim_hawrylak_prb2008}. 
We will call a system of coupled QDs an artificial molecule since the constituent dots are connected 
by the tunnelling which determines the state of the system. 
The presence of the doubly-occupied dot enables us to generate a triplet ground state due to the tunnelling between QDs. 
When all three dots are on resonance, a triple quantum dot system in triangular geometry with four electrons has a spin triplet ground state. 
As we lower the energy level of one of the dots, this dot will have two electrons and the other two dots will have single electron each. 
If the energy level of the doubly occupied dot is well below the energy levels of the other two dots, 
the system is similar to a double quantum dot with spin singlet ground state. 
Therefore, there is a critical value of the bias that changes the ground state between spin singlet and spin triplet. 
Near the critical value, the effective interaction between the two localized electrons 
in the singly occupied dots can be tuned by the bias, both in magnitude and sign. 
When tuned to be in triplet ground state with total spin S=1, 
the TQD molecule can be used as a building block of a spin-one Heisenberg chain system.
The architecture of the proposed linear chain of TQD molecules in a semiconductor device is shown in Fig.~\ref{fig:setting}(a). 
By patterning the top metallic gate in a typical metal-oxide-semiconductor field-effect-transistor (MOSFET) device, 
one can create local potential minima in the two-dimensional electron gas (2DEG) and each potential minimum represents a quantum dot. 
Progress in the experimental technique allows precise control of the potential profile of the 2DEG and the number of electrons 
in each QD \cite{ciorga_sachrajda_prb2000,petta_johnson_science2005,koppens_buizert_nature2006,gaudreau_studenikin_prl2006,gaudreau_kam_apl2009}. 
For individual control of each QD, many plunger gates will be needed 
and the gate pattern would be more complicated than shown in Fig.~\ref{fig:setting}(a). 
Each TQD molecule consists of three quantum dots of which one contains two electrons 
and the other two dots contain single electron [see Fig.~\ref{fig:setting}(b)]. 
The doubly occupied dots in neighbouring TQD molecules are in the opposite position 
to minimize the effects of the Coulomb interaction of those dots. 

The physics of a quantum dot network is well described by the Hubbard model \cite{korkusinski_gimenez_prb2007,delgado_shim_prb2007}. 
For simplicity, we will assume that the on-site Coulomb repulsion $U$ is the same for all QDs, 
the Coulomb interaction between two electrons in different dots in the same molecule is $V$, 
and two electrons in different molecules interact via Coulomb interaction $V'$ only when they are in neighbouring QDs. 
The intra-molecular tunnelling is $t$, and the inter-molecular tunnelling $t'$ is only allowed between two neighbouring dots. 
In an isolated TQD molecule, we can occupy one of the dots with two electrons and the other two dots with single electron 
by lowering the potential energy of the doubly occupied dot (or, equivalently, raising the potential energy of the singly occupied dots). 
In the present TQD chain system, the inter-molecular Coulomb interaction $V'$ 
acts as if it effectively raises the energy levels of the singly occupied dots compared to the doubly occupied dots. 
By controlling $V'$ the ground state of a TQD molecule can be changed from singlet to triplet. 
The only exceptions are two end dots which have no neighbouring molecules. 
Thus we assume that the energy levels of the two end dots are raised by $V'$ through gate control.
In second quantized form, the Hubbard Hamiltonian with single level per site is given by
\beqa
\widehat{H}_{\mathrm{Hubbard}} 
&=& \sum_{i,\sigma} \vep_i c^{\dag}_{i\sigma} c_{i\sigma} 
   +\sum_{i\ne j}\sum_{\sigma} t_{ij} c^{\dag}_{i\sigma} c_{j\sigma} 
   +\sum_i U \hat{n}_{i\uparrow} \hat{n}_{i\downarrow} 
   +\frac{1}{2}\sum_{i\ne j}V_{ij} \hat{n}_{i} \hat{n}_{j}  \nonumber\\
&\equiv& \widehat{H}_0 + \widehat{H}_T + \widehat{H}_U + \widehat{H}_V ~,
\eeqa
where $i,j$ are indices for QD sites, $\sigma=\uparrow,\downarrow$ is the spin, 
$\hat{n}_{i\sigma}=c^{\dag}_{i\sigma} c_{i\sigma}$ is the number of electrons with spin $\sigma$ in QD $i$, 
and $\hat{n}_{i}=\hat{n}_{i\uparrow} + \hat{n}_{i\downarrow} $ is the total number of electrons in QD $i$.
$t_{ij}=t$ and $V_{ij}=V$ for $i$ and $j$ in the same molecule and
$t_{ij}=t'$ and $V_{ij}=V'$ if $i$ and $j$ are in different molecules but neighbouring each other. 
All the other $V_{ij}$ and $t_{ij}$ are zero. The QD energy level is $\vep_i=\vep_0$ for all $i$ 
except for the first and last QDs where $\vep_i=\vep_0+V'$. 
Most of the parameters are highly tunable but, in typical devices, 
the strong Coulomb repulsion is much larger than tunnelling between dots. 
In those cases, the number of electrons in each dot is well defined and 
the tunnelling Hamiltonian $\widehat{H}_T$ gives rise to effective interactions between well localized electrons.
In following numerical examples we use $\vep_0=0$, $U$=2.0, $V$=0.5, $V'$=0.2, $t$=-0.05 and $t'$=-0.02 
in unit of the effective Rydberg defined by $Ry=m^*_e e^4 / 2 \epsilon^2 \hbar^2$.
$m^*_e$ is the electron effective mass, $e$ is the electron charge and $\epsilon$ is the dielectric constant. 
For GaAs, for example, $Ry$ is about 6meV. 
These parameters are based on the first experimental realization of a lateral TQD molecule device \cite{gaudreau_studenikin_prl2006} 
and the theoretical model for the charging diagram of the device \cite{korkusinski_gimenez_prb2007}. 
The sign of the tunnelling elements $t$ and $t'$ can be determined by comparing the Hubbard model 
and more microscopic linear-combination-of-harmonic-oscillator (LCHO) model \cite{korkusinski_gimenez_prb2007,delgado_shim_prb2007}.

\begin{figure}
 \includegraphics[width=0.8\linewidth]{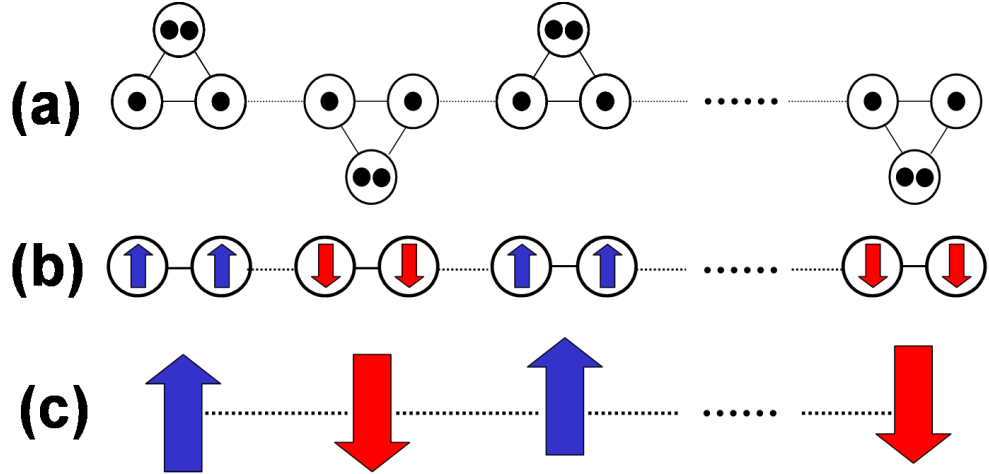}\\
 \caption{(Color online) Different models for the chain of triple quantum dot molecules. 
          (a) In Hubbard model, each dot has a single level and they are connected by the tunnelling and Coulomb interaction. 
          (b) When the Coulomb interaction is strong and the tunnelling can be considered as a perturbation, 
          the singly occupied dots are represented with localized spins with alternating spin-spin interaction. 
          (c) If the intra-molecular ferromagnetic interaction $J$ is much stronger than the inter-molecular antiferromagnetic interaction $J'$, 
          each molecule is in spin triplet state and the system can be considered as a spin-one antiferromagnetic Heisenberg chain.}
 \label{fig:models}
\end{figure}

\begin{figure}
 \includegraphics[width=0.8\linewidth]{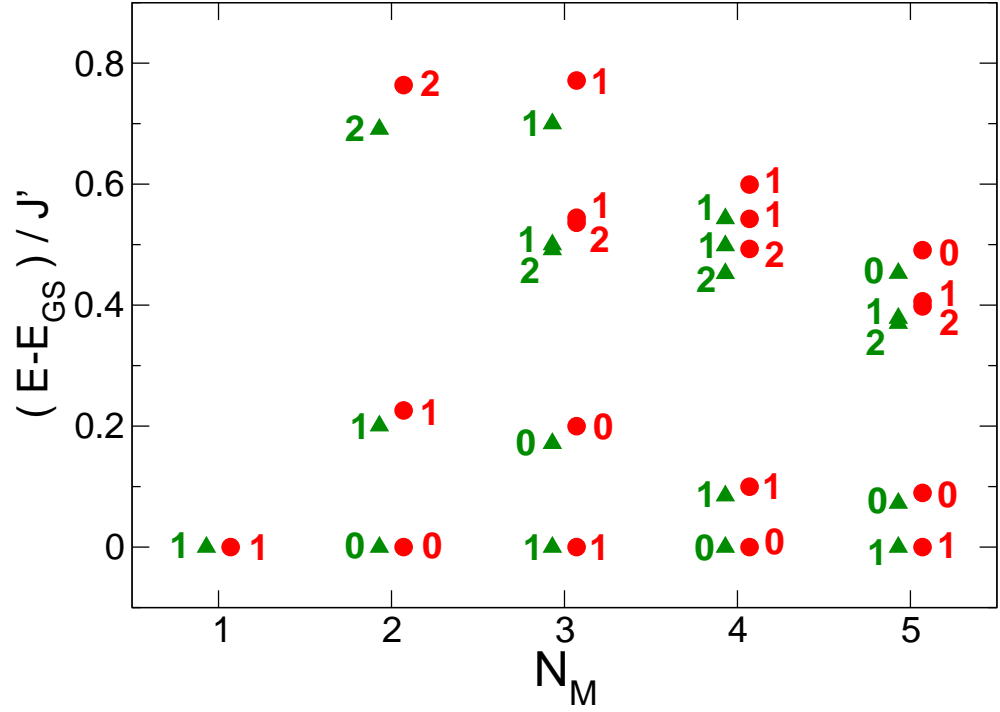}\\
 \caption{(Color online) Comparison of Hubbard model and spin-half chain model. 
          The effective Hamiltonian of spin-half chain with alternating spin-spin interaction describes 
          the low energy states of the full Hubbard Hamiltonian very well both qualitatively and quantitatively. 
          Green triangles are the lowest eigenstates of the Hubbard Hamiltonian and red circles are the eigenstates 
          of the effective spin-half chain Hamiltonian. The numbers adjacent to the triangles and circles represent 
          the total spin of each eigenstates. The result is shown for TQD chains with up to 5 molecules. }
 \label{fig:tqdchain_spinhalf_compare}
\end{figure}

For a system with a charge configuration as shown in Fig.~\ref{fig:models}(a), 
we can obtain an effective Hamiltonian corresponding to a Hamiltonian 
of a chain of localized electron spins [Fig.~\ref{fig:models}(b)] by treating the tunnelling Hamiltonian $\widehat{H}_T$
as a perturbation for the parameter range $|t|,|t'| \ll V,V' \ll U$ \cite{shim_hawrylak_prb2008}. 
The resulting spin-half chain has alternating ferromagnetic and antiferromagnetic interactions 
as envisaged in the works of ,e.g., Hida \cite{hida_prb1992} and Hung and Gong \cite{hung_gong_prb2005}. 
The effective Hamiltonian is given by
\beqa
\widehat{H}_{\mathrm{eff}} 
&=& E_0 + E'_0 
  + J \sum_{m=1}^{N_M} \mathbf{s}_{3m-2}\cdot \mathbf{s}_{3m}  
  + J' \sum_{m=1}^{N_M-1} \mathbf{s}_{3m}\cdot \mathbf{s}_{3m+1} ~,
\eeqa
where
\beqa
E_0 &=& 4N_M\vep_0 + N_M U + 5 N_M V + \left( N_M+1\right)V' ~, \nonumber \\
E'_0 &=& -\frac{2|t|^2 N_M}{V'}-\frac{|t|^2 N_M}{U-V}-\frac{|t'|^2 \left( N_M-1\right)}{U-V'}-\frac{|t|^3 N_M}{V'^2} ~, \nonumber\\
J &=& \frac{4|t|^2}{U-V}-\frac{4|t|^3}{V'^2} ~,\nonumber\\ 
J' &=& \frac{4|t'|^2}{U-V'}~. 
\eeqa
$m$ is the index for TQD molecules and $N_M$ is the number of TQD molecules in the chain. 
The intra-molecular spin-spin interaction $J$ consists of two terms. 
The first term is antiferromagnetic (positive sign) superexchange term and the second term is 
ferromagnetic (negative sign) term which comes from third order processes involving the doubly occupied dot. 
This third order term is comparable to the second order superexchange term in magnitude, 
and for small enough $V'$ the overall interaction $J$ can be either ferromagnetic or antiferromagnetic 
depending on the values of the parameters. Near the critical value of $V'$ where $J=0$, 
each molecule has one doubly occupied dot and two singly occupied dots, 
and the effective spin-spin interaction between the two localized spins changes 
between ferromagnetic and antiferromagnetic. 
The critical value of $V'$ is
\beq
V'_c = \sqrt{|t|(U-V)}~,
\eeq
and we need to tune $V'$ to be smaller (larger) than $V'_c$ for (anti)ferromagnetic intra-molecular interaction. 
On the other hand, the inter-molecular spin-spin interaction between two neighbouring electrons is always 
antiferromagnetic due to the superexchange. 
This effective spin-half chain Hamiltonian gives a clearer picture of the system 
and describes the low energy physics as will be shown below. 
Another important advantage of this effective Hamiltonian is that it reduces the dimension of the Hilbert space 
significantly and allows us to study much longer chains by exactly solving the associated eigenvalue problem. 
The solution of the full Hubbard Hamiltonian is numerically demanding for chains with more than 5 molecules 
since the dimension of the Hilbert space rapidly exceeds $10^6$. 
Figure~\ref{fig:tqdchain_spinhalf_compare} shows the lowest eigenvalues of the Hubbard model (green triangles) 
and the effective spin-half chain model (red circles) for a chain of up to 5 molecules 
using Jacobi-Davidson method implemented in PRIMME library \cite{stathopoulos_mccombs_siamjsc2007}. 
Each eigenstate is labelled by the total spin $S$. 
The values of $J=-5.83\times 10^{-3}$ and $J'=8.89\times 10^{-4}$ used in Eq. (2) 
are obtained from parameters of the microscopic model using Eq. (3). 
It is clear that the effective spin-half chain Hamiltonian describes the low energy levels 
of the Hubbard model quite well, reproducing the main features 
such as (i) the spin singlet ($S=0$) ground state for even number of molecules 
and spin triplet ($S=1$) ground state for odd number of molecules, 
(ii) the decrease of the gap between spin singlet and spin triplet states with the number of molecules, 
and (iii) appearance of the Haldane gap between the ground state and spin quintuplet ($S=2$) excited states.
In the limiting case of $J/J' \rightarrow -\infty$, {\it i.e.}, if the intra-molecular ferromagnetic interaction 
is extremely strong compared to the inter-molecular antiferromagnetic interaction, 
each molecule will be in a triplet state and the system will be a chain of spin-one dimmers [Fig.~\ref{fig:models}(c)]. 
Thus the ground state will be separated from excited states by Haldane gap 
as the length of the chain increases. Even for modest values of $J/J'$, 
the system will have a robust ground state with a finite energy gap. 

\begin{figure}
 \includegraphics[width=0.8\linewidth]{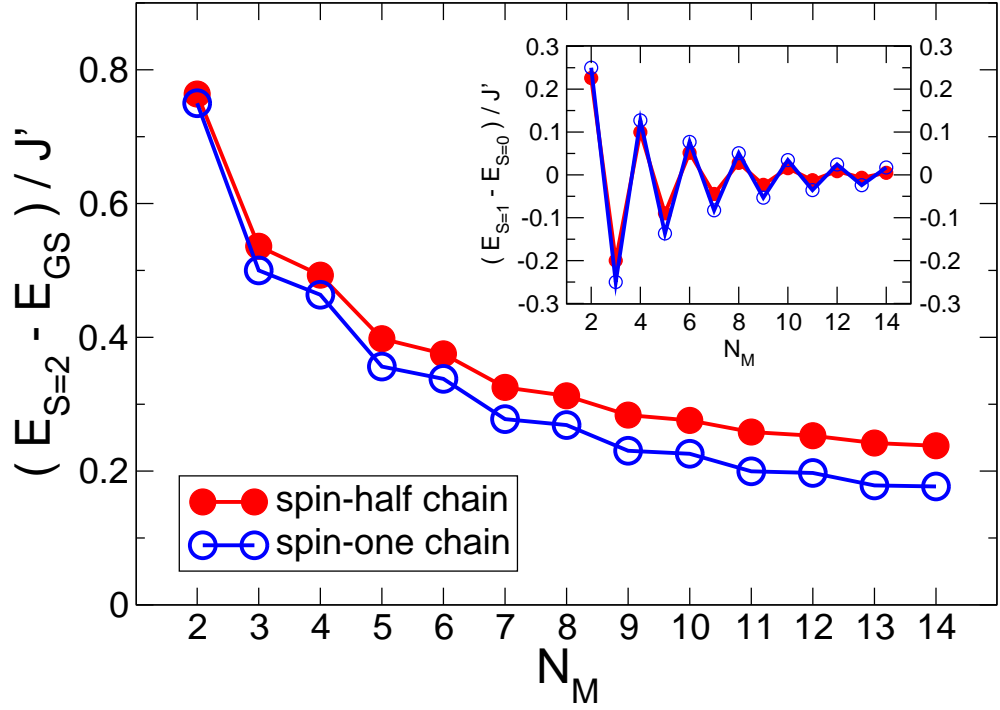}\\
 \caption{(Color online) Comparison between spin-half chain and spin-one chain. 
          The main panel shows the energy gap between spin-quintuplet($S$=2) and ground state, 
          in unit of the effective antiferromagnetic interaction $J'$ of the spin-half chain. 
          The energy gap reaches a finite value at infinite chain, which is the Haldane gap. 
          Spin-half chain has a larger energy gap than the spin-one chain. 
          The inset shows the spin-triplet ($S$=1) and spin-singlet ($S$=0) energy gap. 
          Both spin-half alternating chain and spin-one chain show oscillating spin singlet and spin triplet ground state 
          and the energy gap between spin singlet and triplet states decreases as the length of the chain increases 
          and goes to zero for infinite chain.}
 \label{fig:spinhalf_spinone_compare}
\end{figure}

Figure~\ref{fig:spinhalf_spinone_compare} shows the formation of Haldane gap as function of increasing number $N_M$ 
of triple quantum dot molecules described by the Heisenberg Hamiltonian, Eq. (2). 
The Haldane gap for $N_M$ up to 5 molecules is also shown in Fig.~\ref{fig:tqdchain_spinhalf_compare}. 
Note that the energies in Fig.~\ref{fig:tqdchain_spinhalf_compare} and Fig.~\ref{fig:spinhalf_spinone_compare} are scaled with respect to 
the inter-molecular antiferromagnetic interaction $J'$ of the spin-half chain. 
The results for spin $S=1/2$ finite chain are compared with results for spin $S=1$ chain in Fig.~\ref{fig:spinhalf_spinone_compare}. 
The antiferromagnetic interaction $J_{AF}^{(1)}$ in spin-one chain was chosen to be 1/4 
of the inter-molecular antiferromagnetic interaction $J'$ of the spin-half chain, 
because the interaction is only between two neighbouring spins in the spin-half chain \cite{hung_gong_prb2005}. 
Main panel of Fig.~\ref{fig:spinhalf_spinone_compare} shows the energy eigenvalues of $S=2$ excited state 
with respect to the ground state and the inset shows the energy difference between $S=1$ and $S=0$ states. 
The triplet-singlet gap oscillates around zero 
since the ground state is spin singlet(triplet) for even(odd) number of molecules, 
and it goes to zero as the number of molecules increases. 
For long chains, the ground state is four-fold degenerate ($S=0,1$) 
and there is a finite gap between the ground state and the first excited state ($S=2$). 
The four-fold degeneracy can be manipulated and removed by adding additional QDs 
with localized spin 1/2 electrons at both ends \cite{white_prl1992}. 
The Haldane gap is $0.41 J_{AF}^{(1)}$ for infinitely long spin-one chain \cite{white_prl1992,white_huse_prb1993}, 
and the energy gap of infinite spin-half chain depends on 
both the ferromagnetic and antiferromagnetic interactions $J$ and $J'$ \cite{hida_prb1992}. 
For $J/J' \rightarrow -\infty$, the effective antiferromagnetic interaction $J_{AF}^{(1)}$ 
of the spin-one chain is exactly $J'/4$. 
For finite $J/J'$ the gap is larger than the Haldane gap and increases as the magnitude of  $J/J'$ decreases, 
reaching the maximum value at $J/J'=0$. Around this point, 
the ground state is simply a chain of singlet dimers and the first excited state is "triplet wave" state \cite{hida_prb1992}. 
With the Hubbard parameters used, we get $J/J'=-6.56$.  
As can be seen in Fig.~\ref{fig:spinhalf_spinone_compare}, the energy gap of spin-half chain is larger than that of the spin-one chain 
due to the finite ferromagnetic intra-molecular interaction and it would reach a value 
about twice the Haldane gap of spin-one chain at infinite length \cite{hida_prb1992}. 
Our estimate for material such as GaAs lead to the gap of infinite TQD chain with $J/J'=-6.56$ to be $1.07 \mu eV$,
which corresponds to $T=12.4\mathrm{mK}$. The maximum gap for $J/J'=0$ is $J'$ or 62mK. 
In addition to changing the ratio $J/J'$, one can further increase the gap and hence strengthen 
the robustness of the ground state through controlling the parameters that determine $J'=4t'^2/(U-V')$ 
by building quantum dot networks using self-assembled quantum dots on nanotemplates \cite{reimer_mckinnon_physicaE2008}. 
For characteristic parameters for self-assembled quantum dots  $U$=20meV, $V'$=10meV, $t'$=10meV 
one can reach $J'$~40meV, exceeding room temperature.  
Such a robust ground state could be used as an essential part of a larger quantum circuit, 
extending existing semiconductor technology to quantum applications.

The Authors thank Ian Affleck for discussions. This work was supported by Canadian Institute for Advanced Research and QuantumWorks.




\newpage


\end{document}